\documentclass{article}
\usepackage{amsfonts,amsmath}

\newcommand{\kt}[1]{\ensuremath{|#1\rangle}}

\newcommand{\bk}[2]{\ensuremath {\langle #1|#2 \rangle}}
\newcommand{\kb}[2]{\ensuremath {| #1 \rangle\langle #2|}}
\newcommand{\vp}{{\bf p}}
\newcommand{\vk}{{\bf k}}
\newcommand{\vP}{{\bf P}}

\newcommand{\HS}{{\mathcal{H}}}
\newcommand{\Tr}{\mathrm{Tr}}

\begin{document}
\setlength{\textheight}{8.0truein}    

\thispagestyle{empty}
\setcounter{page}{1}

\vspace*{0.88truein}

\centerline{\bf
Continuous-Discrete Entanglement: An Example with Non-Relativistic Particles
\vspace*{0.035truein}}
\vspace*{0.37truein}
\centerline{\footnotesize
N.L.~Harshman}
\vspace*{0.015truein}
\centerline{\footnotesize\it Department of Computer Science, Audio Technology and Physics, American University}
\baselineskip=10pt
\centerline{\footnotesize\it 4400 Massachusetts Ave., NW}
\baselineskip=10pt
\centerline{\footnotesize\it Washington, DC 20016-8058, United States of America}
\vspace*{10pt}

\vspace*{0.21truein}

\begin{abstract}
This article discusses entanglement between two subsystems, one with discrete degrees of freedom and the other with continuous degrees of freedom.  The overlap integral between continuous variable wave functions emerges as an important parameter to characterize this kind entanglement.  ``Beam-like'' entanglement and ``shape-like'' entanglement are contrasted.  One example of this kind of entanglement is between between the spin degrees of freedom and the momentum degrees of freedom for a non-relativistic particle.  This intraparticle entanglement is Galilean invariant.
\end{abstract}

\vspace*{10pt}

Keywords: continuous variable entanglement, spin-momentum entanglement
\vspace*{3pt}

\vspace*{1pt}   
\section{Introduction}        
\vspace*{-0.5pt}
\noindent
The article considers the properties of entanglement between two subsystems, one of which has discrete degrees of freedom (DOF) and the other has continuous DOF.  This topic therefore lies somewhere between the more well-established field of entanglement in discrete systems~\cite{virmani} and the growing field of continuous variable (CV) entanglement~\cite{eisert_03}.
The subsystems under consideration need not be ``systems'' in the sense of two different quantum objects, like electrons or atoms or fields.  Following recent work~\cite{zanardi01, zanardi04,barnum04}, by the term ``subsystem'' all that is implied is that there is an algebra of observables that can be divided into two commuting subalgebras and this can be used to specify a tensor product structure.

Two recent applications of continuous-discrete (CD) entanglement are in analyses of the Stern-Gerlach experiment~\cite{roston} and nuclear decays~\cite{bertulani}.  In \cite{roston}, the entanglement discussed is between the spin DOF and momentum DOF of a single particle.  In \cite{bertulani}, the entanglement is between the total momentum DOF and the total spin DOF of two particles.  One goal of this work is to put these analyses into the broader context of CD entanglement and identify commonalities.

First, we will consider this problem mathematically, without reference to any particular physical system.  In this context, the importance of the overlap integral between the different CV wave functions will become apparent.  Generally, the smaller the overlap, the greater the entanglement.  There are two fundamentally different ways that small overlap can occur:  (1) The wave functions could have support in different domains of the continuous variables.  We use the description ``beam-like'' for this kind of CD entanglement.  (2) The wave functions could have support in similar domains of the continuous variables, but the details of their structure lead to cancellations.  We call this kind of CD entanglement ``shape-like''. In the case of zero overlap, the different CV wave functions act just like discrete DOF.

As an example, we will consider a form of CD entanglement that can arise in particles: spin-momentum entanglement of a non-relativistic electron.  This kind of intraparticle entanglement is invariant under Galilean transformations between inertial reference frames.  Therefore, it has an intrinsic physical meaning.  In the concluding example, we will exemplify the two different extremes of CD entanglement considering the properties of Gaussian momentum wave functions. 

\section{Continuous-Discrete Entanglement}
\vspace*{-0.5pt}
\noindent
We will consider a system that has a complete set of commuting observables (CSCO) that partitions into three sets.  One set will be the observables \[
\{\hat{P}_1, \hat{P}_2,..., \hat{P}_d\}.
\]
These observables will be represented as operators with a continuous spectrum in the system's Hilbert space representation.  Examples are the momentum generators of the Heisenberg group in $d$-dimensions or the generators of translations in $\mathbb{R}^d$.  Another ``set'' will be the single operator $\hat{S}_i$, which will have a discrete spectrum in the Hilbert space representation.  A simple example of $\hat{S}_i$ is a component of the angular momentum, but this notion could be generalized to other compact Lie algebras or to a set of commuting observables with discrete spectra.  Finally, there are the generalized Casimir observables.  These will be represented as multiples of the unit operator in the system's Hilbert space and although they typically are important for characterizing the physical system, they will play no role in this analysis.

From the existence of this kind of CSCO, it is implied that the system's Hilbert space can be factored as $\HS=\HS_s\otimes\HS_p$.  To the discrete DOFs corresponds the $n$-dimensional Hilbert space $\HS_s = \mathbb{C}^n$.  The Hilbert space associated to the $d$-dimensional continuous DOFs is the space of Lebesgue-integrable functions $\HS_\vp = L^2(\mathbb{R}^d)$, although in practice, we will limit ourselves to a dense ``well-behaved'' subspace $\mathcal{S}_\vp\subset L^2(\mathbb{R}^d)$ (see below).

We can define a basis $\kt{\vp,\chi}=\kt{\chi}\otimes\kt{\vp}$ for $\HS$ in terms of a discrete label $\chi\in\{0,1,...,n-1\}$ and continuous label $\vp\in\mathbb{R}^d$.  These are the eigenvectors of the operators $\{\hat{P}_1,...,\hat{P}_d\}=\hat{\vP}$ and $\hat{S}_i$:
\begin{equation}\label{basis}
\vP\kt{\vp,\chi} = \vp\kt{\vp,\chi},\ \hat{S}_i\kt{\vp,\chi}=((n-1)/2 - \chi)\kt{\vp,\chi}
\end{equation}
The eigenvalues of the $\hat{S}_i$ operator are written as $((n-1)/2 - \chi)$ for later application when it will be interpreted as spin component along some axis.

Because $\hat{\vP}$ is an unbounded operator, the basis eigenkets $\kt{\vp}$ are not elements of $\HS_\vp$ and can only defined as anti-linear functionals on nuclear subspaces of $\HS_\vp$~\cite{bohmrhs}.  One suitable choice for this subspace is the Schwartz space $\mathcal{S}_\vp$ of well-behaved vectors, i.e.\ vectors corresponding to smooth, rapidly-decreasing wave functions.  For vectors in a nuclear subspace $\kt{\phi}\in\HS_s\otimes\mathcal{S}_\vp$, the basis eigenket expansion holds:
\begin{equation}\label{basexp}
\kt{\phi}= \sum_{\chi=1}^n \int d^d\vp \phi_\chi(\vp) \kt{\vp, \chi},
\end{equation}
where the eigenket normalization is
\begin{equation}
\bk{\vp,\chi}{\vp',\chi'} = \delta_{\chi\chi'}\delta^d(\vp -\vp').
\end{equation}
Because of the restriction to the Schwartz space and the normalization, the integration measure $d^d\vp$ can be chosen as the standard Riemann integration measure in $d$-dimensions~\cite{bohmrhs}, e.g. $d^d\vp=\Pi_{i=1}^d dp_i$.

As a consequence of the product structure of the Hilbert space, one can talk about internal entanglement between the discrete and continuous DOFs.  For a pure state $\kt{\phi}$, one can define the reduced discrete density matrix $\rho^s$ by
\begin{eqnarray}
\rho^s &=& \Tr_\vp (\kb{\phi}{\phi})\nonumber\\
&=& \sum_{\chi,\chi'}\left( \int d^d\vp \phi_\chi(\vp)\phi^*_{\chi'}(\vp) \right)\kb{\chi}{\chi'}\nonumber\\
&=& \sum_{\chi,\chi'} h_{\chi,\chi'}\kb{\chi}{\chi'},
\end{eqnarray}
where $h_{\chi,\chi'}$ are the integrals of overlap of CV wave functions and the matrix elements of the reduced discrete density matrix.
Also, one can define the reduced continuous density matrix (operator) $\rho^\vp$ by
\begin{eqnarray}
\rho^\vp &=& \Tr_s (\kb{\phi}{\phi})\nonumber\\
&=& \int d^d\vp d^d\vp'\left( \sum_\chi \phi_\chi(\vp)\phi^*_{\chi}(\vp') \right)\kb{\vp}{\vp'}\nonumber\\
&=& \int d^d\vp d^d\vp' f(\vp,\vp')\kb{\vp}{\vp'}.
\end{eqnarray}
Assuming the initial pure state is normalized,
\begin{equation}
\sum_\chi \int d^d\vp |\phi_\chi(\vp)|^2 = 1,
\end{equation}
then both reduced density matrices are normalized
\begin{equation}
\Tr_\vp \rho^s = \Tr_s \rho^\vp = 1.
\end{equation}
Additionally, for pure states $\kb{\phi}{\phi}$, but not generally for mixed states, we have
\begin{equation}\label{functeq}
\Tr_\vp g(\rho^s) = \Tr_s g(\rho^\vp),
\end{equation}
where $g$ is a continuous, analytic function on the domain of Hermitian operators on both $\HS_s$ and $\mathcal{S}_\vp$.  This is a consequence of the fact that the matrices $\rho^s$ and $\rho^\vp$ have the same spectrum of non-zero eigenvalues~\cite{tommasini98}.

Because finite matrices are easier to work with, it will be easier to consider the reduced discrete density matrix $\rho^s$ for calculating entanglement properties of pure states.
From normalization, the sum of the diagonal elements of the reduced discrete density matrix is unity:
\begin{equation}\label{chinorm}
\sum_\chi h_{\chi,\chi} = 1.
\end{equation}
The off-diagonal elements $h_{\chi,\chi'}=h_{\chi',\chi}^*$ are the integral of overlap between different wave function components $\phi_\chi$ and $\phi_{\chi'}$
\begin{equation}
h_{\chi,\chi'}=\int d^d\vp \phi_\chi(\vp)\phi^*_{\chi'}(\vp).
\end{equation}
The magnitude of these off-diagonal elements are bounded from above by $1/\sqrt{2}$ (for any $n$) by combination the Cauchy-Schwartz inequality
\begin{equation}\label{chiineq}
0\leq |h_{\chi,\chi'}|^2 \leq |h_{\chi,\chi}||h_{\chi',\chi'}|
\end{equation}
and (\ref{chinorm}), which implies that
\begin{equation}
|h_{\chi,\chi}||h_{\chi',\chi'}|\leq 1/2.
\end{equation}

The amount overlap between the wave function components determines how entangled the state is.  To see how this works, consider the extreme cases of maximum overlap and no overlap.
\begin{itemize}
\item \emph{Maximum overlap:}  The first limiting case is when all components of the wave function have the same momentum dependence.  In this case the wave function has the form
\begin{equation}\label{sep}
\phi_\chi(\vp) = c_\chi \phi(\vp)
\end{equation}
where 
\begin{equation}
\sum_\chi |c_\chi|^2 = 1
\end{equation}
and
\begin{equation}
\int d^d\vp |\phi(\vp)|^2 = 1
\end{equation}
The overlap will be truly maximal in the case where $c_\chi = \omega_\chi/\sqrt{n}$, where $|\omega_\chi|=1$. In any case where (\ref{sep}) is valid, the discrete and continuous DOF can be separated and the discrete reduced density matrix elements are
\begin{equation}
h_{\chi,\chi'}=c_\chi c^*_{\chi'}.
\end{equation}
This is the density matrix of a pure state and the entanglement as measured by any entanglement monotone should be zero.

\item \emph{Zero overlap:} The other extreme is when $h_{\chi,\chi'}=0$ unless $\chi=\chi'$.  This in turn breaks into two cases:  (1) All $\phi_\chi(\vp)$ are identically zero except for one.  This state is clearly separable and has no CD entanglement.  (2) The more interesting case is when   $\phi_\chi(\vp)$ and $\phi_\chi' (\vp)$ are orthogonal for all $\chi,\chi'$.  Then the continuous DOFs can effectively be treated as discrete DOFs; and $\chi$ then plays a dual role labeling the orthogonal CV mode and the discrete index.

When $h_{\chi,\chi'}=0$, the discrete density matrix $\rho^s$ is already in diagonalized form, and so the entanglement depends only on the relative magnitudes of the diagonal elements $h_{\chi,\chi}$.  Maximal entanglement occurs when when $\rho^s$ is maximally mixed and therefore $|h_{\chi,\chi}|=1/n$.
\end{itemize}

In the second case of orthogonal wave functions, it may be important to contrast two different kinds of orthogonality.  One kind, which we metaphorically call beam-like, is when the wave functions are orthogonal because they have support on disjoint regions of $\mathbb{R}^d$.  For example, imagine $d$-dimensional Gaussian wave functions whose central maxima are distant compared to their widths, like focused beams going in different directions.  Hyperentangled states~\cite{hyper} and the interferometry states used in \cite{kolar} are examples of these.  Alternatively, one could imagine that the wave functions have the similar domains but are close to orthogonality because of cancellations in the integration, like the various families of orthogonal polynomials.  We call this shape-like CD entanglement.  For example, the $\phi_\chi(\vp)$ wave functions could be energy eigenstates of some potential well expressed in the momentum basis.

\section{Intraparticle Spin-Momentum Entanglement}
\vspace*{-0.5pt}
\noindent
We now consider the non-relativistic particle as an example of CD entanglement.  For non-relativistic particles, the state space of a particle can be factored into the direct product of two Hilbert spaces $\HS_e=\HS_s\otimes\HS_\vp$, where $\HS_s$ is associated to the spin DOF and $\HS_\vp$ to the momentum DOF.  This separation is allowed by the CSCO $\{\hat{\vP},\hat{S}_i; \hat{M}, \hat{W}, \hat{\bf S}^2\}$.  The eigenvalues of the invariant operators for a single free particle $\{\hat{M}, \hat{W}, \hat{\bf S}^2\}$ are the mass $m$, internal energy $W=E - p^2/2m$, and intrinsic spin extracted from $s(s+1)$.  These invariants characterize the representation space $\HS(m,W,s)$, and since we will consider the electron, $s=1/2$ and $m=m_e$ is a known constants, and without loss of generality we can consider the case $W=0$.   All this can be proved by  explicitly constructing the unitary (projective) representation spaces of the (centrally-extended, universal covering of the) Galilean group $\mathcal{G}$ of symmetries~\cite{gal}.  This representation space corresponds precisely to the solution of the two component, non-relativistic Schr\"odinger equation for a free electron.

As described above, we will consider the dense subspace $\HS_{1/2}\otimes\mathcal{S}_\vp$ of the electron Hilbert space $\HS_e = \HS(m_e,0,1/2)$.  Then the momentum eigenkets will be well-defined and we will have the usual expansion theorems: every ``well-behaved'' pure state vector in $\kt{\phi}\in\HS_s\otimes\mathcal{S}$ and can be expanded as
\begin{eqnarray}\label{exp}
\kt{\phi}&=& \sum_\chi \int d^3\vp \phi_\chi(\vp) \kt{\vp, \chi}\nonumber\\
&=& \int d^3\vp \sum_\chi \phi_\chi(\vp) \kt{\vp, \chi}.
\end{eqnarray}
We now interpret $\hat{\vP}$ as a momentum operator and $\hat{S}_i$ as the spin component in some direction.  
By choosing the Schwartz space $\mathcal{S}_\vp$ for the dense subspace, we guarantee that the expectation values of all positive powers of momentum are bounded, as well as other properties assuring that the momentum wave functions are ``well-behaved''~\cite{bohmrhs}.  This restriction makes physical sense and the importance of energy boundedness for entanglement measures of CV systems has been previously emphasized~\cite{eisert_02}.

Note that the order of the integral and sum is not relevant in (\ref{exp}) because the Hilbert space of the particle factors into independent DOFs.  This is not the case for relativistic particles because for them the spin component definition is momentum dependent~\cite{czachor}.

Because of the product structure of the nonrelativistic, single particle Hilbert space, one can talk about the spin-momentum entanglement of a particle in a frame-independent way.  All shifts of reference frame are Galilean transformations and these are represented as a tensor product of unitary operators acting locally on each factor Hilbert space.  Here we show the explicit construction of the unitary irreducible representations of the Galilean group in the representation that is implied by our choice of CSCO:

A general element $g\in\mathcal{G}$ is denoted by ten parameters: $g=(b,\mathbf{a}, \mathbf{v}, R)$, where $b$ is the amount of time translation, $\mathbf{a}$ is the three-vector of space translation, $\mathbf{v}$ is the three-vector of velocity boost, and $R\in\mathrm{SO}(3)$ is a rotation.  The projective representation acting on the basis for the physical subspace of $\HS_e$ can be chosen as
\begin{equation}
U(g)\kt{\vp, \chi} = e^{i(\frac{1}{2}m_e \mathbf{a}\cdot\mathbf{v} - \mathbf{a}\cdot\mathbf{\vp} + bE)}\sum_{\chi'}D^{1/2}(R)_{\chi\chi'}\kt{\vp', \chi'},
\end{equation}
where $\vp' = R\vp + m\mathbf{v}$ and $D^{1/2}(R)$ is a $2\times 2$ unitary matrix representing $R$ in $\HS_{1/2}$ (some details of the $\mathrm{SU}(2)$-$\mathrm{SO}(3)$ little group homomorphism have been swept under the table here).    Notice, as promised, this representation factorizes into unitary operators on the discrete Hilbert space, i.e.\ the unitary matrix $D^{1/2}(R)$, and the continuous Hilbert space, i.e.\ the phase factor.  Since Galilean transformations are tensor products of local unitaries with respect to the spin-momentum tensor product structure, any reasonable measure of entanglement will be invariant across different inertial reference frames.  Note, however, that in the relativistic case, the spin-momentum tensor product structure is not invariant under  Poincar\'e symmetry, as a result, ``mixing'' between different kinds of entanglement can occur~\cite{czachor, relent} (although sometimes authors have not recognized the true origin of this phenomenon).

The reduced spin density matrix $\rho^s$ of the pure electron state $\kt{\phi}$ will have the following form
\begin{equation}
\rho^s = \left( \begin{array}{cc} h_{0,0} & h_{0,1} \\ h_{0,1}^* & h_{1,1} \end{array}\right).
\end{equation}
The eigenvalues $\lambda_\pm$ of this matrix characterize the entanglement properties:
\begin{equation}
\lambda_{\pm} = \frac{1}{2} \pm \sqrt{\frac{1}{4} - (h_{0,0}h_{1,1}- |h_{0,1}|^2)}.
\end{equation}
For maximum entanglement, $\lambda_{\pm}=1/2$.  From (\ref{chiineq}), the quantity $h_{0,0}h_{1,1}- |h_{0,1}|^2$ is always positive and less than $1/4$.  The upper bound corresponds to maximum entanglement and the lower bound to separability.

Separability will occur in two cases: (1) Either $\phi_0(\vp)$ or $\phi_1(\vp)$ is identically zero.  The state is obviously separable then because the electron is polarized in the direction of one spin component.  (2) In the other case, $\phi_0(\vp)=\omega\phi_1(\vp)$ where $|\omega|=1$.  Then $h_{0,0}h_{1,1}=|h_{0,1}|^2$ and the state can be put in the form (\ref{sep}) where $\omega$ is absorbed in $c_0$.  All other $\kt{\phi}$ will be entangled, and this entanglement increases as $\phi_0(\vp)$ and $\phi_1(\vp)$ come closer to equal norm of $1/\sqrt{2}$ and to orthogonality.

\section{Concluding Example}
\vspace*{-0.5pt}
\noindent
Let us now consider a specific but flexible model for considering beam-like entanglement.  Each wave function component $\phi_\chi(\vp)$ will be taken to be a Gaussian in 3D momentum space with central momentum  $\vk_\chi$ and width $\sigma_\chi$:
\begin{equation}\label{model}
\phi_\chi(\vp) = \frac{c_\chi}{(\pi\sigma_\chi^2)^{3/4}}\exp\left(-\frac{1}{2\sigma_\chi^2}(\vp-\vk_\chi)^2\right).
\end{equation}
The constant $c_\chi$ gives the normalization of each Gaussian wave function, i.e.
\begin{equation}
\int d^d\vp |\phi_\chi(\vp)|^2 = |c_\chi|^2
\end{equation}
and total normalization implies $|c_0|^2 + |c_1|^2=1$.
These kinds of states could be useful approximations for wave packets prepared under some dynamics that have an explicit asymmetry with respect to a preferred projection of spin component, the classic example being the Stern-Gerlach device~\cite{bertulani}.

The reduced spin density matrix for the non-relativistic electron with wave functions of the form (\ref{model}) is
\begin{equation}
\rho^s = \left( \begin{array}{cc} |c_0|^2 & c_0c_1^*x \\ c_0^*c_1x^* & |c_1|^2 \end{array}\right)
\end{equation}
where $x$ is the Gaussian overlap function
\begin{eqnarray}
x &=& \int d^3\vp \phi_0(\vp)\phi^*_1(\vp)\nonumber\\
&=& \left(\frac{2\sigma_0\sigma_1}{\sigma_0^2 + \sigma_1^2}\right)^{3/2}\exp\left(\frac{-2q^2}{\sigma_0^2 + \sigma_1^2}\right)
\end{eqnarray}
and $q = |\vk_0 - \vk_1|$ is the magnitude of the difference of the central momenta.
The eigenvalues of this matrix are
\begin{equation}\label{schmidt}
\lambda_{\pm} = \frac{1}{2} \pm \sqrt{\frac{1}{4} - |c_0|^2|c_1|^2(1 - |x|^2)}.
\end{equation}

The first observation from (\ref{schmidt}) is that maximum intraparticle entanglement can only occur if the state has a spin component expectation value $\langle \hat{S}_i \rangle =0$.  For such states, the quantity $|c_0|^2|c_1|^2$ takes its maximal value $1/4$.  This feature is a consequence of the intrinsic preferred direction in the choice of states (\ref{model}), and is not a general feature of intraparticle entanglement.  

The second observation is that intraparticle entanglement increases when $|x|\rightarrow 0$, i.e.\ when the Gaussian wave packets have minimal overlap.  This can be seen most easily in the case where $|c_0|^2|c_1|^2=1/4$; then (\ref{schmidt}) simplifies to
\begin{equation}
\lambda_{\pm} = \frac{1}{2} \pm \frac{1}{2} |x|.
\end{equation}
The quantity $|x|$ will go to zero if either or both $2q^2 \gg \sigma_0^2 + \sigma_1^2$ and $\sigma_0/\sigma_1 \gg 1$ (or equivalently $\sigma_0/\sigma_1 \ll 1$).  Physically, these limits correspond to narrow wave packets with different central momenta or to wave packets with same central momentum and very different widths, respectively.  Both cases can best be thought of as examples of beam-like entanglement.

Beam-like entanglement is manifest in particle experiments with beam splitters, e.g. the Stern-Gerlach experiment, and in quantum optics with photon.  It is unclear whether it is experimentally feasible to produce shape-like entanglement in a free, beam-like system, but perhaps the control of wave functions possible within quantum optics could allow such a possibility.  In bound electron systems, shape-like entanglement between vibrational energy levels and spin states has already be achieved with trapped ions~\cite{benkish}.  The orthogonality of the bound states allows the CV DOF to be treated like discrete DOF.

section*{Acknowledgements}
\noindent
The author would like to acknowledge the U.S.-Italian Fulbright Commission and the University of Trento for supporting this work during a sabbatical from American University  Additional support for the research was provided by the the Research Corporation.  Also, the author is grateful to the organizers and participants of the 38th Symposium on Mathematical Physics in Toru\~n, Poland in June 2006 for an excellent conference and fruitful discussions, and especially to Eirik Ovrum and Konrad Banaszek.

\end{document}